\documentclass[aps,prl,onecolumn,12 pt,superscriptaddress,showpacs]{revtex4}
\usepackage{amssymb}
\usepackage{graphicx}

\begin{document}

\title{Persistence of characteristics of an ordered flux line lattice above the second peak in $Bi_2Sr_2CaCu_2O_{8+ \delta}$.}
\author{A. Pautrat}\author{Ch. Simon}\author{C. Goupil}
\affiliation{CRISMAT/ENSI-Caen, UMR 6508 du CNRS, 6 Bd Marechal
Juin, 14050 Caen, France.}
\author{P. Mathieu}
\affiliation{Laboratoire Pierre Aigrain de l'Ecole Normale Sup\'erieure, UMR 8551 du
CNRS, associ\'ee aux universit\'es Paris 6 et 7, 75231 Paris Cedex5, France.}
\author{A.\ Br\^{u}let}
\affiliation{Laboratoire L\'{e}on Brillouin,  CEN Saclay, 91191
Gif/Yvette, France.}
\author{C. D. Dewhurst}
\affiliation{Institut Laue Langevin, 6 rue Jules Horowitz,
Grenoble, France.}
\author{A.I. Rykov}
\affiliation{Siberian Synchrotron Radiation Centre, Novosibirsk,
Russia.}

\begin{abstract}

We report Small Angle Neutron Scattering measurements of
the flux lines lattice (FLL) in $Bi_2Sr_2CaCu_2O_{8+ \delta}$. As previously reported, the scattered intensity decreases strongly when the 
magnetic field is increased, but it remains measurable far above the
second peak. The direct observation of Bragg peaks proves that the characteristics of a lattice
are still present. No structural features related to a symmetry breaking, such as 
a liquid like or an amorphous state, can be observed. However, the associated
 scattered intensity is very low and is difficult to explain. We discuss the coexistence between two FLL states as a possible interpretation.
\end{abstract}

\newpage
\maketitle
\section{introduction}
Since the discovery of high temperature superconductors,
a lot of work has been done to understand their transport and magnetic
properties  \cite{giam}. In general, these properties are closely connected with the pinning mechanisms of the flux lines lattice (FLL).
 If bulk pinning is significant,
a disordering of the FLL can cause an increase
of the pinning efficiency. The sharp second peak in the critical current (or in the magnetization)
 is then usually associated with a transition between a FLL and a phase without long range order (a vortex glass) \cite{giam}.
 It is therefore important to
obtain clear information on the correlation between the FLL order and the pinning properties. Small angle neutron
scattering (SANS) is the dedicated technique, which allows to probe the
FLL in the bulk of a sample, by observation of Bragg peaks. Several groups have observed
that, in high $\kappa$ superconductors, the Bragg peaks associated
 with FLL order are observed for small magnetic fields but disappear quickly when this field is increased
\cite{bob,klein1,klein2}. The disappearance of the intensity is generally attributed to a strong disordering of
the FLL. Different interpretations have been proposed: a
dimensional cross-over in $Bi_2Sr_2CaCu_2O_{8+ \delta}$ ($Bi-2212$)
\cite{bob}, a Debye Waller-like effect in $BaKBiO_{3}$
\cite{klein1}. When the intensity decreases, no broadening of the Bragg peaks seems to take place. This important result
 may indicate a moderate
increase of the effective disorder in a weak decay correlation
function. This behavior is consistent with the
Bragg-Glass phase \cite{klein2}. In this scenario , the
intensity drop toward zero reflects the dislocations proliferation
leading the destruction of the long range ordering (Bragg-Glass
melting). This transition from a FLL (Bragg Glass) towards a disordered state would be the transition responsible for the second peak \cite{giam,klein2}. 
 It remains however one point
 which seems essential to clarify: the fact that the intensity is zero (below the experimental resolution) prevents any analysis of the presumedly disordered state.
 Consequently, the signatures of diffraction of a glass or a liquid (a ring of
scattering and broadened Bragg peaks, analog to a powder or liquid
diffraction pattern) which are necessary to conclude on the nature of the state were not observed \cite{bob,klein2}.
 Finally, there is no direct proof of the nature of the FLL state in the second peak. 
Note also that a strong decrease of the scattered intensity has been recently reported in $LaSr_{1.9}Sr_{0.1}CuO_{4}$, but at a field rather different
 from the one of the second peak \cite{divakar}. This suggests that the
scenario of a second peak caused by the destruction of FLL order may be not systematically relevant.

\section{experimental}
We present here a study of the intensity scattered from the
FLL in $Bi-2212$. Our sample is a monolithic single crystal of $Bi-2212$ (30$\times$5$\times$1.2 $mm^3$), oriented with $(110)^*$ along
 the length and $c^*$ parallel to both the magnetic field and
 the neutron beam. The crystal is very slightly overdoped with $T_{c} \approx$ 87 $K$. In the literature, we find that
 typical parameters for this doping range are $\lambda_{ab}(0) =$ 260
 $nm$ \cite{aegerter} and $\xi_{ab} (0) =$ 1.5 $nm$ \cite{li}, giving a high Ginzburg-Landau parameter $\kappa = \frac{\lambda_{ab}}{\xi_{ab}} \approx$ 173.
  The value of $\xi_{ab}$ remains largely dubious, since it is deduced by extrapolating measurements made at high temperature.
From the point of view of the nuclear structure, the typical mosaic spread was $\Delta
\theta \lesssim$ 1 $deg$ \cite{Notabene}, what is quite reasonable
for such a large crystal. Our experiment was carried out on the
D22 SANS instrument, at the Institut Laue Langevin (France). The incident
neutron wavelengths used were 9 $\AA$ and 15 $\AA$, with a
resolution of $\Delta\lambda / \lambda\thickapprox$ 10 $^{\circ}
/_{\circ}$. The scattered intensity was recorded on a 2D
multidetector (128$\times$128 pixels) located at 17.6 $m$ from the
sample position. The data were taken at $T=$ 4.2 $K$ with applied
fields 0.01 $T \lesssim  B \lesssim$ 0.1 $T$ after field cooling
 the sample from 90 $K (> T_c)$. Due to the
substantial small angle scattering background, all presented data
are differences between the raw data and the background taken at
zero field at $4.2$ K, or at fixed field above $T_c$.

\section{results and discussion}
In order to correlate the structural characteristics of the FLL with the magnetic properties, 
we cleaved a small piece of sample from the crystal which was used for SANS. It was then studied in a SQUID magnetometer.
 To estimate the position of the second peak, we took the same criterion which is used in the literature \cite{aegerter,bernard}.
This is the
point of maximum slope of the magnetisation $M(B)$, just
before the maximum of $M$. One finds here that the field of second peak is $B^{*} \approx$ 0.04 $T$ (see fig.1a). This is in good agreement with other
reports \cite{aegerter}. In fig.1a, typical SANS patterns are shown for different values of the magnetic field.
Each image is the sum of scattering from the FLL, as the sample is rocked horizontally and vertically through the Bragg condition (30 $+$ 30 discrete angular settings). 
We find that the diffracted intensity is centered with the value of $Q_{10} \simeq  2\pi/(1.07\sqrt{\phi_0/B})$ for all the values of magnetic
 field which we studied (0.01 $T$-0.1 $T$). This corresponds to the vector of diffraction of the hexagonal FLL. 
The integrated intensity $I_{hk}$, for a $(hk)$ Bragg reflection, normalized to the neutron flux is given by \cite{thorel}:
\begin{equation}
I_{hk} =  2 \pi(\frac{\gamma}{4})^2 V (\frac{\lambda_n}{\phi_0})^2 \frac{|F_{hk}|^{2}}{Q_{hk}}
\end{equation}
where $F_{hk}$ is the form factor for the field distribution within one FLL unit cell,
 $\lambda_n$ is the neutron wavelength, $V$ is the sample volume, $\gamma$= 1.91 is the
 gyromagnetic ratio of the neutron.
 It is also convenient to define the reflectivity $R=I_{hk}/S$ where $S$ is the illuminated sample surface \cite{morten}.
 As shown in fig. 1b, the reflectivity can be measured up to $B =$ 0.1 $T > B^{*}$ with reasonable counting times.
 It decreases notably with the magnetic field, and seems to decrease more quickly for $B \geq  B^*$. Anyway, no
first order structural transition, which should be marked by a
sharp collapse of the intensity, can be depicted at $B = B^*$. Note that a similar intensity
 decrease has been observed in $Nd_{1.85}Ce_{0.15}CuO_{4}$.
 This has been attributed to a cross over to a more disordered state \cite{gilardi}.
Clearly, some intensity remains, even if it is weak, for fields appreciably higher than $B^*$.
 If one put aside this strong decrease of intensity to focus only in the structural signatures,
 the FLL seems non disturbed while crossing $B^*$. Note that in \cite{bob,klein1,klein2} the FLL signal was observed to disappear at $B^*$.
  We think that this was essentially due to a limited resolution. The fact that we observe here intensity above this field is due to several factors: the very large size of our sample, the high neutrons flux of D22 at the ILL.
We have also notably increased the counting times and we have changed of the neutron wavelength to optimize the ratio signal over noise for measuring the very low intensities.

 The intensity as a function of the azimuthal angle and the 
rocking curves can be fitted with Gaussian or Lorentzian curves. The azimuthal width of the peaks $\Delta \psi$ (Fig.2a)
 and the rocking curves width $\Delta \phi$ (Fig.2b) does not change significantly up to $B = 0.08 T$.
This means that a well
ordered FLL, without degraded orientational order, persists
notably above $B^{*}$.
 One can notice that the angle $\psi$ tends to increase for the strongest fields. 
 However, this increase
 is not really significant because of the large error bars due to the very low intensities (Fig.2a). We can conclude
 this part by the persistence of FLL order
when it crosses the field of the second peak $B^ *$. This is the most important result of this study.

In fig.3, we show the form factor $F_{10}$ which has been deduced from the equation (1). To analyze this form factor,
 we can start with the simplest model which is the London limit (point
core size approximation and linear superposition). It should be valid in our case where case $b= B/B_{c2} \ll 1$ and $\kappa \gg 1$ \cite{notebene2,Brandta}.
The London form factor is \cite{mumut2}:

\begin{equation}
F_{hk}=\frac{B}{1+\lambda^{2}Q_{hk}^{2}}
\end{equation}

where $\lambda$ is the London length. At $B$= 0.01 $T$,
 one finds $\lambda = 220 \pm 20 nm$, in reasonable agreement with values deduced from previous SANS
 or $\mu SR$ experiments \cite{bob,aegerter}. The scattered intensity was enough to measure
second order Bragg reflections. As it was interestingly shown by \textit{Kealey et al} in
$Sr_2Ru_2O_4$ \cite{kealey}, the ratio $F_{11}/F_{10}$ can be used as a good test of the validity of the London (or of another) model.
 Using (1) and (2) and $\lambda =$ 220 $nm$, one calculates $F_{11}/F_{10}= 0.389$ for B= $0.01$ T,
similar to the experimental value $0.36 \pm 0.015$. This shows that the London model seems to be a reasonable starting point for the weakest measured field.
However, one can easily realize that this expression leads to a very weak dependence of the form factor with the field whereas a strong fall is observed (Fig.4).
 This fact was even the principal experimental support to propose a FLL transition induced by the magnetic field \cite{bob,klein1,klein2}.
 To explain the decrease of $F_{10}$ in the framework of the London model, a field dependent $\lambda$ can be involved, due for example to $d_{x^2-y^2}$ pairing \cite{sauls}.
 Nevertheless, the magnitude of the observed effect seems really too large for this interpretation.
In the following, we will let $\lambda$ as a constant and discuss the variation of $F_{10}$ with a functional form which is different from the London one.  

In such a restricted range of data
 (here about one decade of magnetic fields, see Fig.3), it is always difficult to make a good choice for this functional form. It can be estimated that 
$F_{10}$ follows a quasi-exponential field dependence with a change of slope at $B \approx $ 0.05 $T$, i.e. close to $B^{*}$ (see Fig. 3).
We emphasize that many other functions can be proposed to fit the data. For example, as one can easily realize by looking at the fig. 3, a Gaussian variation of $F_{10}$ as a function of $B$ (or $Q^2$)
could be a good choice. Nevertheless, we do not know any theoretical justification for such a dependence.

On the contrary, there are at least two ways to obtain an exponential dependence of the factor of form. The first one is to correct the unphysical 
 point core size of the London model by adding a cut-off factor. A Gaussian cut-off  $exp(-Q^2 2\xi^2)$ can be used at low field \cite{Brandta}.
Note that the corrections given by the Clem analytic model \cite{clem}, or by taking into account more properly the finite
 size of vortex core \cite{Brandta}, appear to be negligible in our experimental situation where $b=B/B_{c2} \ll 1$  (typically, the corrections are in (1-b). Here, $B \leq $ 0.1 $T$ and $B_{c2}$ is though to be in the 100 $T$ range). 
The other possibility to have this kind of dependence is to introduce a Debye-Waller (DW) effect. The DW effect, applied to the FLL case, could come from uncorrelated distortion due to the
static disorder \cite{klein1,mcpaul}. The form factor is multiplied by $exp (- Q^2 \langle U^2 \rangle /4)$ \cite{variance}.
 $\langle U^2 \rangle$ is the root mean square displacement around
the equilibrium position $a_0$.

Even if the underlying physics is it very different, each correction leads to:

\begin{equation}
F_{hk}=\frac{B}{1+\lambda^{2}Q_{hk}^{2}} exp (- Q_{hk}^2 \alpha )
\end{equation}

with $\alpha = \langle U^2 \rangle /4$ (DW corrections) or $\alpha = 2\xi^{* 2}$ (Gaussian cut-off).

One of the consequences of this correction is to strongly attenuate the high order Fourier components.
Taking the experimental $F_{10}$ and using equation (3) with $\lambda =$ 220 $nm$, one can extract the value
 of $\alpha$ ($\alpha\approx$ $162 nm^2$ for $B \leq$  0.05 $T$).
 Then, it is possible to calculate the expected $F_{11}/F_{10}$.
In our experiment, when $B>$ 0.05 $T$,
the intensity of $F_{11}$ is below our resolution.
However, at least in the restricted field range available, both experimental and calculated $F_{11}/F_{10}$ are comparable (see Fig.4), what shows some consistencies to use the equation (3).
Consequently, we will discuss the variation of $F_{10}$ as a function of the field using equation (3). 

One can first try to analyse the decrease of intensity with a pure DW effect \cite{klein1}.
The parameter deduced from the experimental $F_{10}$ is then $\langle U^2 \rangle / {a_0^2}$. The Lindeman criterion gives the maximum value of uncorrelated displacements, before 
 that the long range ordering breaks. This criterion is essentially phenomenological, but
much of the theoretical work assumes or concludes that the criterion is $\sqrt{\frac{\langle U^2 \rangle}{a_0^2}} \lesssim 0.25$ \cite{lindeman}.
 For example, free-energy functional for the dislocation density was applied for the FLL case and gives a value of 0.2 \cite{jan}.
Our measurements lead to a much higher value. One finds $\langle U^2 \rangle / {a_0^2} \geq 0.5$ at 0.1 $T$, 
giving the result that a lattice exists for unexpected large mean square displacements. This means that such a Lindeman analysis has to be taken
 with some caution. 
 Note that the intensity decrease without apparent broadening of the Bragg peaks is also compatible with a Bragg Glass phase. This was deeply discussed in \cite{klein2}.
 A modeling of this decrease with the classical elastic theory predicts $I.Q_{10}/F_{London}^2 \propto B^{-2}$, i.e. a rather smooth
 power-law. Such a dependence does not allow to describe our data. Finally, it seems that DW corrections or the existence of a Bragg-Glass phase do not offer a \textit{quantitative} explanation of the intensity decrease in $Bi-2212$.  

If one tries to analyse the data using the Gaussian cut-off, 
the relevant parameter is $\xi^*$. $\xi^*$ is an effective core size,
 and does not necessary reflect the true orbital coherence length $\xi_0=\sqrt{\phi_0 / 2 \pi B_{c2}}$. The Clem-model
 gives $\xi^* \approx \sqrt(2) \xi_0$ \cite{Brandta,clem}, so $\xi^*$ is expected to be reasonably close to $\xi_0$.
The bulk value of $\xi_0$
deep in the superconducting state of $Bi-2212$ is actually unknown, but, from extrapolation, is thought to be in the range of $1-2$ nm.
If we estimate that the low field and high field regime shown in Fig.3 corresponds to two different effective $\xi^*$, we calculate  $\xi^{* }\approx$ 9 $nm$ for $B\leq  0.05T$ and 25 $nm$ for $B\geq 0.05T$.
 At least the second value seems much too large to have its original meaning of an effective core size.

Finally, it seems difficult to explain the low values of scattered intensity by using reasonable parameters. 
This intensity is very low but Bragg peaks are still observed and thus the FLL order is preserved. 
Note also that this result does not seem very consistent with the $\mu SR$ experiments which
 show that the distribution of the magnetic field strongly changes at $B^*$ \cite{lee}.
 
Since the whole of the results is difficult to understand if a homogeneous system is considered, a coexistence between two FLL states can be proposed.
 The first state is the traditional FLL. Its quantity decreases with the magnetic field,
 without any other structural disturbance (there is no broadening of the Bragg peaks),
 and this reduction is directly related to the measured intensity. Note that in such an experiment, we are sensitive to the intensity centered on the Bragg condition (here $\pm$ 1.5 degree).
 If the second FLL state is very disordered,
 the rocking curve is very broad and can contribute almost like a background noise. 
 Another possibility is that the Bragg angle of the second FLL is different, for example because of a rotation, 
 even weak, from the direction of the applied magnetic field. In both cases, the field distribution inside the sample should be inhomogeneous
 at the sample scale what can be consistent with the $\mu SR$ experiments. 
Our present data do not allow a relevant analysis to differentiate or even validate these assumptions. 
We expect, thanks to future SANS experiments, to be more conclusive on the validity of these assumptions.

A coexistence between two FLL states after a field 
cooling makes think of the peak effect in $NbSe_2$.
In this latter case, the FLL state at low field has been clearly identified as a conventional FLL (or a Bragg Glass) \cite{Nb}.
 The structure of the high field state is less clear.
 A disordered state has not been confirmed by experiments measuring the FLL order.
 Decoration experiments have shown that no amorphous state is present in the peak effect region of $NbSe_2$ \cite{fasano}.
 SANS measurements have shown that the FLL state obtained after field cooling,
 does not appear particularly disordered in the bulk but rather turned from the magnetic field direction.
 A peculiar distribution of surface currents has been proposed \cite{Nb}. 
 Transport experiments suggest that these latter can be also important in $Bi-2212$,
 and metastable transport properties, very similar properties with those observed in $NbSe_2$, are also observed \cite{BI}.
It is very important to note that if there is a coexistence between two states, the states observed here under the strongest magnetic fields can be metastable.
 In such a case, their relative quantities can be modified by making various magnetic or thermal histories.
 The intensity measured would be also modified. 
 We note that metastability was already observed in $Nb$ showing the peak effect \cite{brown}. Since this latter seems to be related to the proximity of surface supraconductivity \cite{brown},
the role of the surface currents to stabilize the field cooled state can certainly not be neglected.

In conclusion, SANS measurements bring a new light on the FLL behavior near the second peak in
$Bi-2212$. We observe a strong and quasi exponential
decrease of the intensity as function of the magnetic field. 
Bragg peaks are observed beyond the field generally associated with
the destruction of the FLL, showing the persistence of FLL order. However, the scattered intensity is very low and
 is difficult to understand quantitatively. A coexistence between two FLL states
 could be an interesting possibility to explain this result, but it must be confirmed by other SANS measurements.
Finally, even if a complete understanding of the FLL behavior in
$Bi-2212$ is still challenging, we hope that these new measurements will allow to clarify the physics which is behind the peak effect.

Acknowledgments: We thank E.H. Brandt (EMP Stuttgart) 
for the circular cell program, and S.Lee (university of St andrews)
 for nicely communicating on the $\mu SR$ results of his group.


\newpage

\vskip 1 cm

\begin{figure}[tbp]
\centering\includegraphics*[width=11cm]{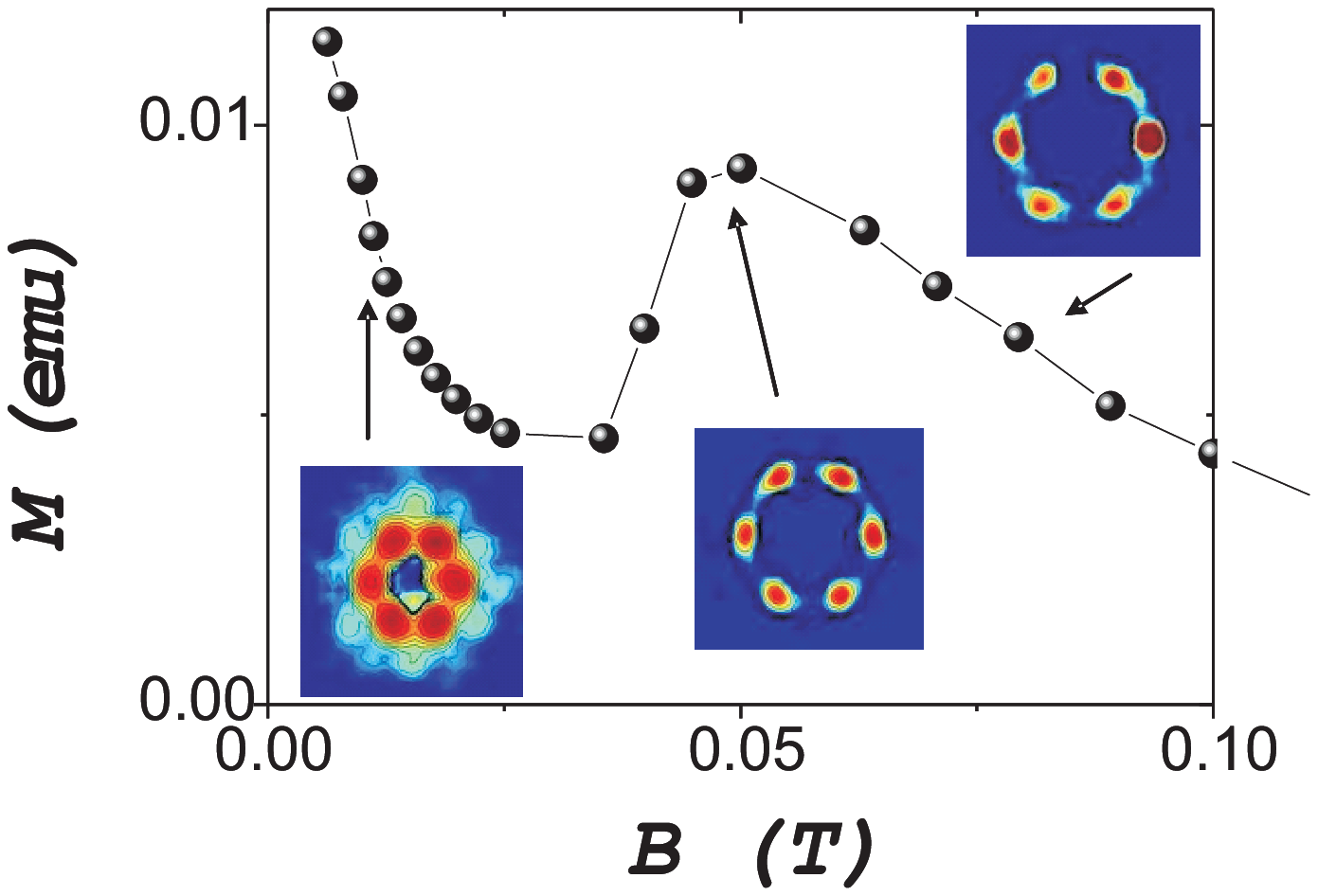}
\vskip 2 cm
\centering\includegraphics*[width=10cm]{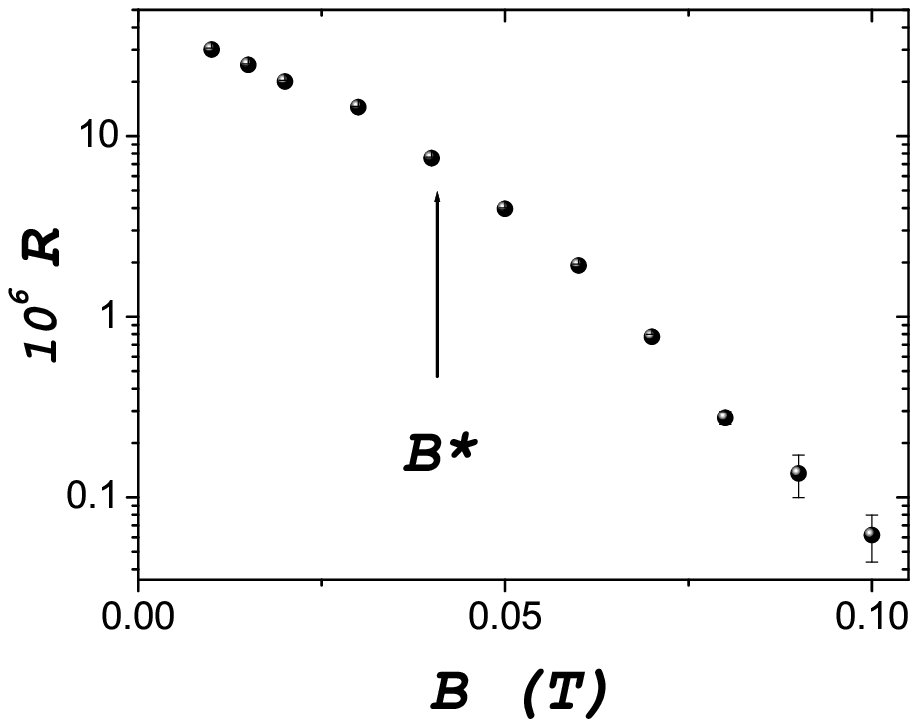}
\vskip 2 cm
 \caption{Color online Top: a/ Low
temperature magnetic hysteresis $\Delta M$ of the $Bi-2212$ sample. The second peak is at
$B^{*} \approx 0.04 T$. Bottom: b/ Reflectivity of the first order Bragg peak as
function of the magnetic field. Note the impressive decrease of the intensity, but without any collapse
when crossing $B^*$. Nevertheless, a change of slope seems to occur close to $B^*$. \label{fig.1}}
\end{figure}

\vskip 1 cm

\begin{figure}[tbp]
\centering\includegraphics*[width=10cm]{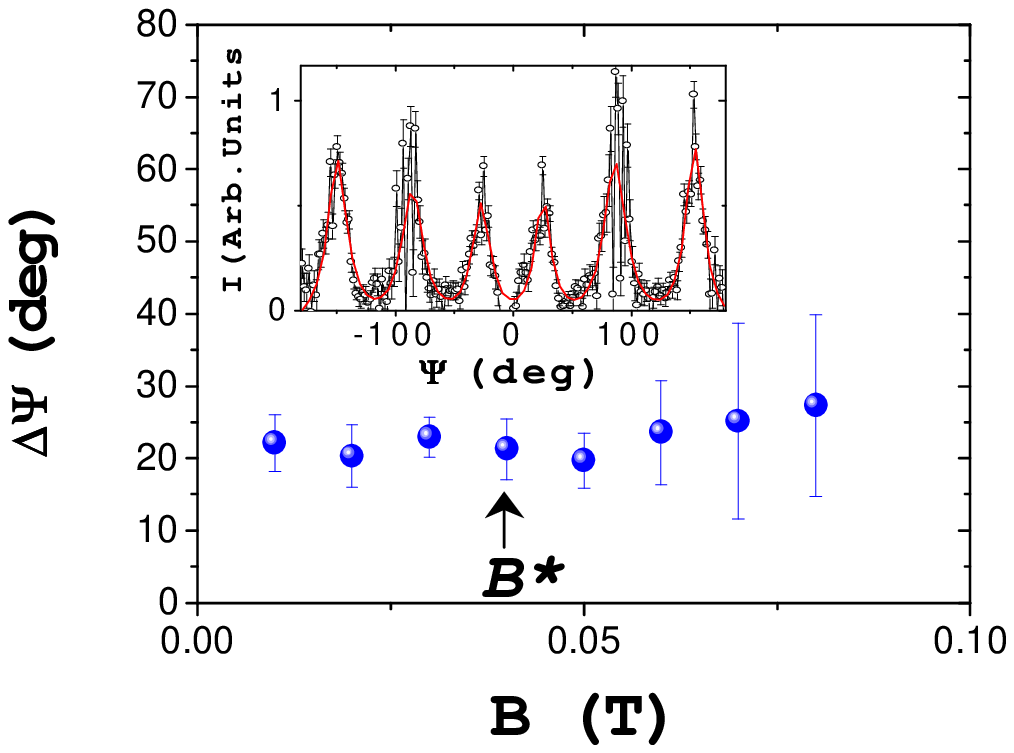}
\vskip 1 cm
\centering \includegraphics*[width=10cm]{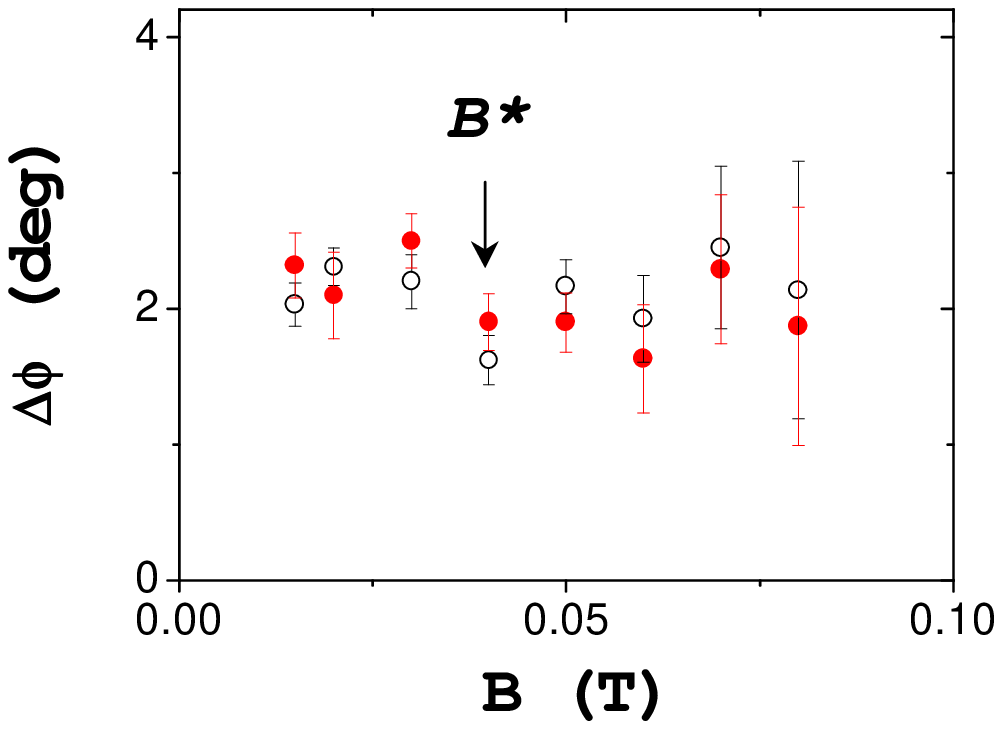}
 \caption{Color online Top: a/ Main azimuthal width as function of the magnetic field (In the inset is
shown the intensity as function of the azimuthal angle for $B =
0.05 T$  proving a robust orientational order even at the maximum of the magnetization peak. Color online Bottom: b/ rocking
curve width as function of the magnetic field. \label{fig.2}}
\end{figure}

\vskip 1 cm

\begin{figure}[tbp]
\centering \includegraphics*[width=10cm]{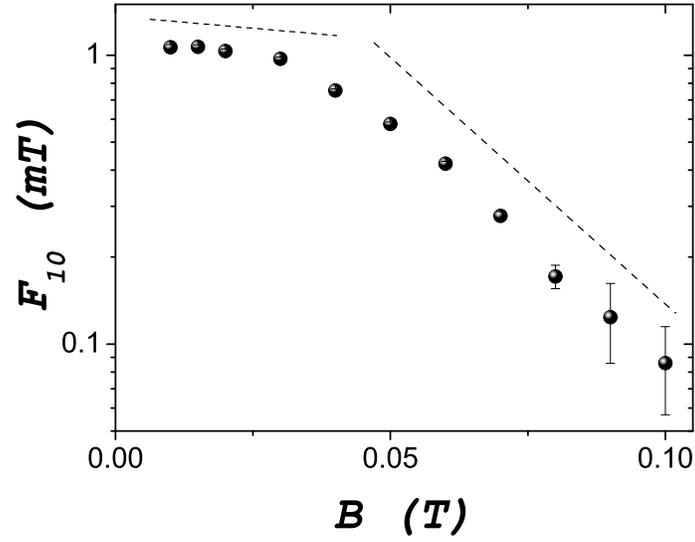}
\caption{$F_{10}$, deduced from fig.1b, as function of the magnetic field. The dashed lines are guides for the eyes. \label{fig.3}}
\end{figure}

\begin{figure}[tbp]
\centering \includegraphics*[width=10cm]{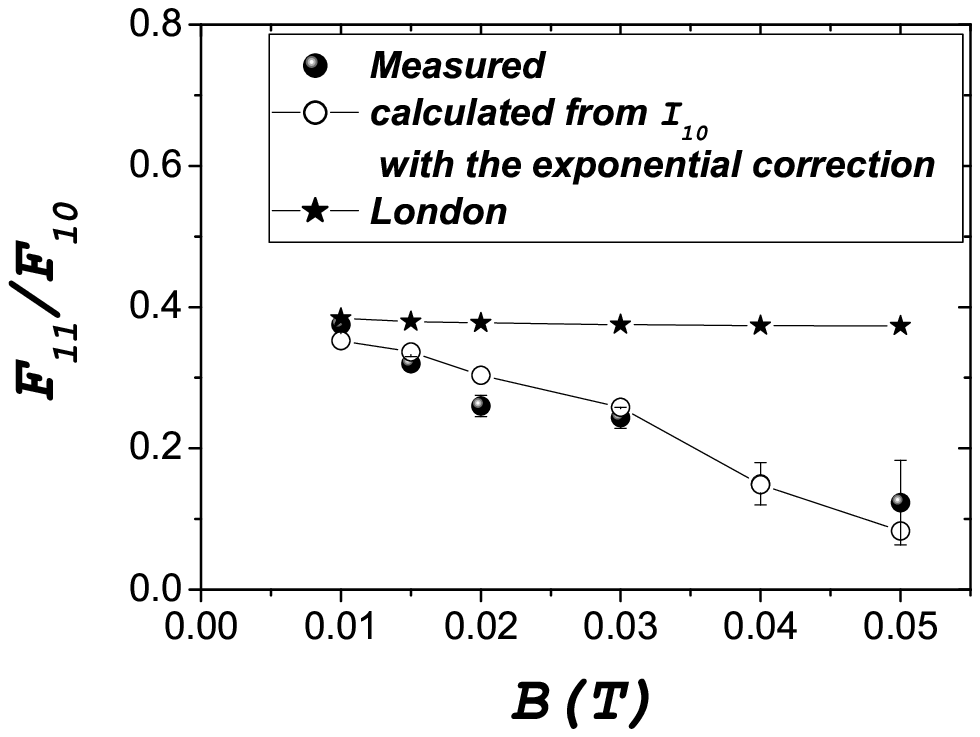} 
\caption{$F_{11}/F_{10}$ as function of the
magnetic field (plain points: experimental data, empty points: $F_{10}$ is calculated from the first order spots intensity using
 equation (3), stars: calculated from the London model with $\lambda$= 220 $nm$ (equation (2)).  
\label{fig.4}}
\end{figure}


\begin{references}
\label{sec:TeXbooks}


\bibitem{giam} T. Giamarchi , S. Bhattacharya, Vortex Phases in ``High Magnetic Fields:
  Applications in Condensed Matter Physics and Spectroscopy", C. Berthier et al., 9, 314,
  Springer-Verlag (2002).


\bibitem{bob} R. Cubitt, E. M. Forgan, G. Yang, S. L. Lee, D. McK. Paul, N. A. Mook, M. Yethiraj, P. H. Kes, T. W. Li, Z. Tarnawski, and K. Mortensen, Nature (London) 365, 407
(1993).

\bibitem{klein1} I. Joumard, J. Marcus, T. Klein, and R. Cubitt, Phys. Rev. Lett. 82, 4930 (1999).

\bibitem{klein2} T. Klein, I. Joumard, S. Blanchard, J. Marcus, R. Cubitt, T. Giamarchi and P. Le Doussal, Nature 413, 406 (2001).

\bibitem{divakar} U. Divakar, A. J. Drew, S. L. Lee, R. Gilardi, J. Mesot, F. Y. Ogrin, D. Charalambous, E. M. Forgan, G. I. Menon, N. Momono, M. Oda, C. D. Dewhurst, and C.
Baines, Phys. Rev. Lett. 92, 237004 (2004).

\bibitem{aegerter} C. M. Aegerter, S. L. Lee, H. Keller, E. M. Forgan, and S. H.
Lloyd, Phys. Rev. B 54, R15661 (1996).

\bibitem{li} M. Li, C. J. van der Beek, M. Konczykowski, A. A. Menovsky,
and P. H. Kes, Phys. Rev. B 66, 024502 (2002).

\bibitem{Notabene} Philippe Bourges (LLB Saclay), private communication. The same sample was used for INS measurements: see L. Capogna, B. Fauqué, Y. Sidis, C. Ulrich, P. Bourges, S. Pailhès, A. Ivanov, J. L. Tallon, B. Liang, C. T. Lin, A. I. Rykov, and B. Keimer
Phys. Rev. B 75, 060502 (2007). 

\bibitem{bernard}C. Bernhard, C. Wenger, Ch. Niedermayer, D. M. Pooke, J. L. Tallon, Y. Kotaka, J. Shimoyama, K. Kishio, D. R. Noakes, C. E. Stronach, T. Sembiring, and E. J. Ansaldo
Phys. Rev. B 52, R7050 (1995).

\bibitem{thorel} P.Thorel, Ph.D. Thesis, Universit$\acute{e}$ de Paris Orsay (1972).

\bibitem{morten} M.R. Eskildsen, Ph.D. thesis, Ris$\oslash$-R-1084(EN), Ris$\oslash$ National
Laboratory, Denmark (1998).(URL address: http://www.risoe.dk/rispub/FYS/ris-r-1084.htm)

\bibitem{gilardi} R. Gilardi, J. Mesot, S. P. Brown, E. M. Forgan, A. Drew, S. L. Lee, R. Cubitt, C. D. Dewhurst, T. Uefuji, and K. Yamada
Phys. Rev. Lett. 93, 217001 (2004).


\bibitem{notebene2}
 Note nevertheless that when $b \approx \kappa^{-2}$, the field profile changes rapidly from a strongly modulated to a quasi-constant variation.
 One should avoid 
approximation and employ exact solutions (E. H. Brandt, Phys. Rev. B 18, 6022 (1978), E. H. Brandt
Phys. Rev. Lett. 78, 2208 (1997)). 
 Unfortunately, the calculation of Fourier components from the Ginzburg Landau equations was not possible due to the huge memory allocation 
needed for the iterative method ($b \ll 1$, $\kappa \gg 1$ is the most unfavourable case). Some modifications are in progress to try to solve this problem.

\bibitem{Brandta} A. Yaouanc, P. Dalmas de Réotier, and E. H. Brandt
Phys. Rev. B 55, 11107 (1997).

\bibitem{mumut2} E. H. Brandt, Phys. Rev. B 18, 6022 (1978).

\bibitem{kealey} P. G. Kealey, T. M. Riseman, E. M. Forgan, L. M. Galvin, A. P. Mackenzie, S. L. Lee, D. McK. Paul, R. Cubitt, D. F. Agterberg, R. Heeb, Z. Q. Mao, and Y.
Maeno, Phys. Rev. Lett. 84, 6094 (2000).

\bibitem{sauls} S.K. Yip and J.A. Sauls, Phys. Rev. Lett. 69, 2264 (1992).

\bibitem{clem} J.R. Clem, J. Low Temp. Phys. 18, 427 (1975). 

\bibitem{mcpaul} M. Yethiraj, D. McK. Paul, C. V. Tomy, and E. M. Forgan,
Phys. Rev. Lett. 78, 4849 (1997).

\bibitem{variance} E.H. Brandt, Phys. Rev. Lett. 66, 3213 (1991).

\bibitem{lindeman} J. Kierfeld and V. Vinokur, Phys. Rev. B 69, 24
501 (2004).

\bibitem{jan} J. Kierfeld and V. Vinokur, Phys. Rev. B 61, R14 928 (2000).

\bibitem{lee} S. L. Lee, P. Zimmermann, H. Keller, M. Warden, I. M. Savic, R. Schauwecker, D. Zech, R. Cubitt, E. M. Forgan, P. H. Kes, T. W. Li, A. A. Menovsky, and Z.
Tarnawski, Phys. Rev. Lett. 71, 3862 (1993).

\bibitem{fasano} Y. Fasano, M. Menghini, F. de la Cruz, Y. Paltiel, Y. Myasoedov, E. Zeldov, M. J. Higgins, and S. Bhattacharya
Phys. Rev. B 66, 020512 (2002).

\bibitem{Nb} A. Pautrat, J. Scola, Ch. Simon, P. Mathieu, A. Br$\hat{u}$let, C. Goupil, M. J. Higgins, and S. Bhattacharya, Phys. Rev. B
71, 064517 (2005). Experiments with a higher neutron flux are now in progress to decide on the genuine nature of the FLL in the peak regime in NbSe$_2$.

\bibitem{BI} A. Pautrat, Ch. Simon, J. Scola, C. Goupil, A. Ruyter, L. Ammor, P. Thopart and D. Plessis, Eur. Phys. J. B 43, 39 (2005).

\bibitem{brown} S. R. Park, S. M. Choi, D. C. Dender, J. W. Lynn, X. S. Ling, Phys. Rev. Lett. Vol.91, 167003 (2003).

\end{references}
\end{document}